\documentclass[superscriptaddress,aps,preprintnumbers,amsmath,showpacs,amssymb,prd,nofootinbib,reprint]{revtex4-1}
\usepackage{amsfonts}
\usepackage[mathscr]{eucal}
\usepackage{dcolumn}
\usepackage{bm}
\usepackage[colorlinks=true,linkcolor=blue,citecolor=blue]{hyperref}
\usepackage{hyperref}
\usepackage{color}
\usepackage{amsmath}
\usepackage{graphics}
\usepackage[all,cmtip]{xy}
\usepackage{float}
\usepackage{longtable}

\usepackage{tikz}
\usepackage{multirow}

\usetikzlibrary{arrows}
\tikzset{>=angle 60}
\tikzstyle{W}=[draw,circle,scale=.6]
\tikzstyle{B}=[draw,circle,fill=black,scale=.6]
\tikzstyle{H}=[draw,circle,fill=gray,scale=.6]
\tikzstyle{every picture}=[scale=.6,baseline=(current bounding box.south)]

\begin{document}


\def\a{\alpha}
\def\b{\beta}
\def\c{\varepsilon}
\def\d{\delta}
\def\f{\phi}
\def\g{\gamma}
\def\h{\theta}
\def\k{\kappa}
\def\l{\lambda}
\def\m{\mu}
\def\n{\nu}
\def\p{\psi}
\def\q{\partial}
\def\r{\rho}
\def\s{\varphi}
\def\t{\tau}
\def\u{\upsilon}
\def\v{\varphi}
\def\w{\omega}
\def\x{\xi}
\def\y{\eta}
\def\z{\zeta}
\def\D{\Delta}
\def\G{\Gamma}
\def\H{\Theta}
\def\L{\zeta}
\def\F{\Phi}
\def\P{\Psi}
\def\S{\Sigma}

\def\aa{{\dot \a}}
\def\bb{{\dot \b}}
\def\ss{{\bar \s}}
\def\hh{{\bar \h}}
\def\CA{{\cal A}}
\def\CB{{\cal B}}
\def\CC{{\cal C}}
\def\CD{{\cal D}}
\def\CE{{\cal E}}
\def\CG{{\cal G}}
\def\CH{{\cal H}}
\def\CI{{\cal I}}
\def\CK{{\cal K}}
\def\CL{{\cal L}}
\def\CR{{\cal R}}
\def\CM{{\cal M}}
\def\CN{{\cal N}}
\def\CO{{\cal O}}
\def\CP{{\cal P}}
\def\CQ{{\cal Q}}
\def\CS{{\cal S}}
\def\CT{{\cal T}}
\def\CW{{\cal W}}

\newcommand{\Slash}[1]{{\ooalign{\hfil/\hfil\crcr$#1$}}}

\def\o{\over}
\newcommand{\gsim}{ \mathop{}_{\textstyle \sim}^{\textstyle >} }
\newcommand{\lsim}{ \mathop{}_{\textstyle \sim}^{\textstyle <} }
\newcommand{\vev}[1]{ \left\langle {#1} \right\rangle }
\newcommand{\bra}[1]{ \langle {#1} | }
\newcommand{\ket}[1]{ | {#1} \rangle }
\newcommand{\EV}{ {\rm eV} }
\newcommand{\KEV}{ {\rm keV} }
\newcommand{\MEV}{ {\rm MeV} }
\newcommand{\GEV}{ {\rm GeV} }
\newcommand{\TEV}{ {\rm TeV} }
\def\diag{\mathop{\rm diag}\nolimits}
\def\Spin{\mathop{\rm Spin}}
\def\SO{\mathop{\rm SO}}
\def\O{\mathop{\rm O}}
\def\SU{\mathop{\rm SU}}
\def\U{\mathrm{U}}
\def\Sp{\mathop{\rm Sp}}
\def\USp{\mathop{\rm USp}}
\def\SL{\mathop{\rm SL}}
\def\tr{\mathop{\rm tr}}
\def\rank{\mathop{\rm rank}}

\def\spin{\mathfrak{ spin}}
\def\so{\mathfrak{so}}
\def\o{\mathfrak{o}}
\def\su{\mathfrak{su}}
\def\u{\mathfrak{u}}
\def\sp{\mathfrak{sp}}
\def\usp{\mathfrak{usp}}
\def\sl{\mathfrak{sl}}
\def\e{\mathfrak{e}}

\def\beq#1\eeq{\begin{align}#1\end{align}}
\def\alert#1{{\color{red}[#1]}}


\preprint{
}

\title{
Chiral algebra of Argyres-Douglas theory from M5 brane
}

\author{
Dan Xie
}
\affiliation{Center of Mathematical Sciences and Applications, Harvard University, Cambridge, 02138, USA}
\affiliation{Jefferson Physical Laboratory, Harvard University, Cambridge, MA 02138, USA}

\author{
Wenbin Yan
}
\affiliation{Center of Mathematical Sciences and Applications, Harvard University, Cambridge, 02138, USA}
\affiliation{Jefferson Physical Laboratory, Harvard University, Cambridge, MA 02138, USA}

\author{
Shing-Tung Yau
}
\affiliation{Center of Mathematical Sciences and Applications, Harvard University, Cambridge, 02138, USA}
\affiliation{Jefferson Physical Laboratory, Harvard University, Cambridge, MA 02138, USA}
\affiliation{Department of Mathematics, Harvard University, Cambridge, MA 02138, USA}


\begin{abstract}
We study chiral algebras associated with Argyres-Douglas theories engineered from M5 brane. For the theory engineered using 6d $(2,0)$ type $J$ theory on 
a sphere with a single irregular singularity (without mass parameter),  its chiral algebra is the minimal model of W algebra of $J$ type.
For the theory engineered using an irregular singularity and a regular full singularity,  its chiral algebra 
is the affine Kac-Moody algebra of $J$ type. We can obtain the Schur index of these theories by computing the vacua character of the corresponding chiral algebra.

\end{abstract}

\maketitle


\section{Introduction
\label{sec:introduction}}
In the past few years, it has been found that certain observables of
four dimensional $\mathcal{N}=2$ superconformal field theories (SCFTs) can be identified with
observables of two dimensional conformal field theories.
Those 4d/2d correspondences include the match between the 4d
sphere partition function and the correlator of 2d Liouville theory \cite{Alday:2009aq}, the match between
the trace of 4d quantum monodromy and the character of certain module of
2d chiral algebra \cite{Cecotti:2010fi,Iqbal:2012xm,Cordova:2015nma,Cecotti:2015lab}. Recently a remarkable map between 4d $\CN=2$ SCFTs and 2d chiral algebras was constructed in \cite{Beem:2013sza}.
In many cases 2d chiral algebras constructed from 4d SCFTs are identified with known 2d models, see
\cite{Gaiotto:2011xs,Beem:2013sza,Beem:2014rza, Buican:2015ina,Song:2015wta,Buican:2016arp,Nishinaka:2016hbw,Cecotti:2015lab}. 

The purpose of this paper is to identify the chiral algebras corresponding to Argyres-Douglas (AD) theories \cite{Argyres:1995jj} engineered from M5 branes \cite{Gaiotto:2009hg,Bonelli:2011aa,Xie:2012hs, Wang:2015mra}.
The basic 4d/2d mappings used in this paper are  \cite{Beem:2013sza}\footnote{A very novel twisting procedure is needed to get those 4d/2d mappings \cite{Beem:2013sza}.}:
\begin{itemize}
\item  The 2d central charge $c_{2d}$ and the level of affine Kac-Moody algebra $k_{2d}$ are related to the 4d central charge $c_{4d}$ and the flavor central charge $k_F$ as
\begin{equation}
c_{2d}=-12 c_{4d},~~k_{2d}=-k_F\footnote{Our normalization of $k_F$ is half of that of \cite{Beem:2013sza,Beem:2014rza}.}.
\label{eq:centralchargerelation}
\end{equation}
\item The (normalized) vacuum character of 2d chiral algebra is the 4d Schur index $\CI(q)$.
\end{itemize}

We consider two types of AD theories engineered from M5 brane and propose that their 
chiral algebras take the following form:
\begin{itemize}
\item The first class $J^b[k]$ is engineered using
a single irregular singularity, and we consider those theories without any flavor symmetry. The chiral algebra is conjectured to be the following coset
\begin{equation}
{\cal A}={J_l\bigoplus J_1 \over J_{l+1}},~~l=-{kh-b\over k},
\end{equation}
with $h$ being the dual Coxeter number of Lie algebra $J$. This is the minimal model of W algebra of type J \cite{Bouwknegt:1992wg}: $W^J(p',p)=W^J(b+k,b)$. 
\item The second class $(J^b[k],F)$ is engineered using a single irregular singularity (no mass parameters) and a full regular singularity. The chiral algebra is conjectured to be
the affine Kac-Moody  algebra:
\begin{equation}
{\cal A}=J_{-k_F},~~k_F=h-{b\over b+k}.
\end{equation}
\end{itemize}

\section{AD theory without flavor symmetry}

Let us start with the theory engineered using one irregular singularity. The irregular singularity has been classified in \cite{Wang:2015mra} and
the corresponding theory is called $J^{b}[k]$,  here $J$ denotes the type of six dimensional $(2,0)$ theory and $b$ is a number specifying types.
The classification is achieved using the classification of irregular solution to Hitchin's equation:
\begin{equation}
\Phi={T\over z^{2+k/b}}+\cdots.
\end{equation}
The corresponding theory can also be realized by putting type IIB string theory on a three-fold singularity whose
form associated with the $J^{b}[k]$ theory is listed  in table \ref{table:isolatedsingularitiesALEfib}.
\begin{table}[!htb]
\begin{center}
  \begin{tabular}{ |c|c|c|c| }
    \hline
       $J$& Singularity & $b$ &$\mu_0$  \\ \hline
     $A_{N-1}$ &$x_1^2+x_2^2+x_3^N+z^k=0$&  $N$& $(N-1)(k-1)$\\ \hline
 $~$& $x_1^2+x_2^2+x_3^N+x_3 z^k=0$ & $N-1$&N(k-1)+1\\ \hline

 $D_N$   & $x_1^2+x_2^{N-1}+x_2x_3^2+z^k=0$ & $2N-2$&N(k-1) \\     \hline
  $~$   &$x_1^2+x_2^{N-1}+x_2x_3^2+z^k x_3=0$& $N$&2k(N-1)-N \\     \hline

  $E_6$  & $x_1^2+x_2^3+x_3^4+z^k=0$&12 &6k-6  \\     \hline
   $~$  & $x_1^2+x_2^3+x_3^4+z^k x_3=0$ &9&8k-6   \\     \hline
  $~$  & $x_1^2+x_2^3+x_3^4+z^k x_2=0$  &8&9k-6   \\     \hline

   $E_7$  & $x_1^2+x_2^3+x_2x_3^3+z^k=0$& 18 &7k-7  \\     \hline
      $~$  & $x_1^2+x_2^3+x_2x_3^3+z^kx_3=0$ &14&9k-7    \\     \hline

    $E_8$   & $x_1^2+x_2^3+x_3^5+z^k=0$&30 &8k-8  \\     \hline
        $~$   & $x_1^2+x_2^3+x_3^5+z^k x_3=0$ &24& 10k-8   \\     \hline
    $~$   & $x_1^2+x_2^3+x_3^5+z^k x_2=0$ & 20 &12k-8 \\     \hline
  \end{tabular}
  \end{center}
  \caption{3-fold singularities corresponding to our irregular punctures\cite{Wang:2015mra}, where $\mu_0$ is the dimension of charge lattice. When $b=h$ with h the dual Coxeter number, such 
  theories are also called $(J, A_{k-1})$  which were first studied in \cite{Cecotti:2010fi}.  }
  \label{table:isolatedsingularitiesALEfib}
\end{table}
\begin{table}[!hb]
\begin{center}
\begin{tabular}{|c|c|c|c|}
\hline
  ${\cal T}$ &$~$&${\cal T}$&$~$  \\ \hline
     $A_{N-1}^N[k]$ &$(k,N)=1$& $A_{N-1}^{N-1}[k]$ &$\text{No solution}$\\ \hline
          $D_{N}^{2N-2}[k]$ &$k\neq 2n$& $D_{N}^{N}[k]$&$N=2^m(2i+1),k\neq 2^m n$\\ \hline
     $E_{6}^{12}[k]$ &$k\neq 3n$& $E_6^{9}[k]$ &$k\neq 9n$\\ \hline
     $E_{6}^8[k]$ &$\text{No solution}$& $E_{7}^{18}[k]$ &$k\neq 2n$\\ \hline
     $E_7^{14}[k]$ &$k\neq 2n,n>1$& $E_{8}^{30}[k]$ &$k\neq 30n$\\ \hline
     $E_{8}^{24}[k]$ &$k\neq 24n$& $E_{8}^{20}[k]$ &$k\neq 20 n$\\ \hline
\end{tabular}
\end{center}
\caption{Constraint on $k$ so that $J^b[k]$ has no flavor symmetry.}
  \label{table:constraint}
\end{table}
We are interested in cases where there are no flavor symmetry. This would put constraints on integer $k$ in each class, see table \ref{table:constraint}.
\begin{table}[!hb]
\begin{center}
\begin{tabular}{|c|c|c|c|}
\hline
  ${\cal T}$ &$c_{4d}$&${\cal T}$&$c_{4d}$  \\ \hline
     $A_{N-1}^N[k]$ &${(N-1)(k-1)(N+k+Nk)\over 12(N+k)}$& $A_{N-1}^{N-1}[k]$ &${(Nk-N+1)(N+k+Nk-1)\over 12(N-1+k)}$\\ \hline
          $D_{N}^{2N-2}[k]$ &${N(k-1)(-2-k+2N+2kN)\over 12(-2+k+2N)}$& $D_{N}^{N}[k]$ &${((N-1)2k-N)(N+k(2N-1))\over 12(k+N)}$\\ \hline
     $E_{6}^{12}[k]$ &${(k-1)(12+13k)\over 2(12+k)}$& $E_6^{9}[k]$ &${(4k-3)(13k+9)\over 6(9+k)}$\\ \hline
     $E_{6}^8[k]$ &${(3k-2)(13k+8)\over4(8+k)}$& $E_{7}^{18}[k]$ &${7(k-1)(19k+18)\over 12(18+k)}$\\ \hline
     $E_7^{14}[k]$ &${(9k-7)(19k+14)\over 12(14+k)}$& $E_{8}^{30}[k]$ &${2(k-1)(30+31k)\over 3(30+k)}$\\ \hline
     $E_{8}^{24}[k]$ &${(5k-4)(24+31k)\over 6(24+k)}$& $E_{8}^{20}[k]$ &${(3k-2)(20+31k)\over3(20+k)}$\\ \hline
\end{tabular}
\end{center}
\caption{4d central charges of theory $J^b[k]$ without flavor symmetry.}
  \label{table:centralcharge1}
\end{table}
The three-fold singularity has a manifest $C^*$ action
\begin{equation}
f(\lambda^{q_i}z_i)=\lambda f(z_i).
\end{equation}
This $C^*$ action is related to the $U(1)_R$ symmetry of $\mathcal{N}=2$ SCFT. The Seiberg-Witten curves for these theories can be
found from the miniversal deformation of the singularity \cite{Shapere:1999xr, Xie:2015rpa}.

The 4d central charge can be computed using the following formula \cite{Shapere:2008zf}:
\begin{equation}
\label{eq:acformula}
a={R(A)\over 4}+{R(B)\over6}+{5 r\over 24},~~c={R(B)\over 3}+{r\over 6},
\end{equation}
where $r$ is the rank of the Coulomb branch, and for our theories \cite{Xie:2015rpa}:
\begin{equation}
R(A)=\sum_{[u_i]>1}([u_i]-1),~~R(B)={1\over 4}\mu u_{max},
\end{equation}
in which $\mu$ is the dimension of the charge lattice and $u_{max}$ is the maximal scaling dimension of Coulomb branch operators, both of which can be expressed in terms of the weights $q_i$'s:
\begin{equation}
\begin{split}
\mu=\prod ({1\over q_i}-1),~~
u_{max}={1\over \sum q_i-1}.
\end{split}
\end{equation}
Other quantities appearing in the central charge formula are also easy to compute
\begin{equation}
r={\mu_0\over 2},~~\mu=\mu_0.
\end{equation}
We list $\mu_0$  in table \ref{table:isolatedsingularitiesALEfib} and central charge $c_{4d}$  in table \ref{table:centralcharge1}.

\subsection{2d chiral algebra}

It was realized in \cite{Cordova:2015nma} that the chiral algebra of $A_{N-1}^{N}[k]$  is  $W^{A_{N-1}}(N+k,N)$ minimal model. What is really important to us is that such model
can be realized as the coset \cite{Bais:1987zk}:
\begin{equation}
{\cal A}={SU(N)_l\bigoplus SU(N)_1 \over SU(N)_{l+1}},
\end{equation}
here $l=-{(k-1)N\over k}$.  Motivated by this coset realization of chiral algebra, we would like to conjecture that the 2d chiral algebra for all our model $J^b[k]$
can be realized as the following diagonal coset model
\begin{equation}
{\cal A}={\frak{g}_l\bigoplus \frak{g}_1 \over \frak{g}_{l+1}},~~\frak{g}=J,~~l=-{hk-b\over k}.
\end{equation}
Here $h$ is the dual Coxeter number. 
The 2d central charge of our coset model is
\begin{equation}
c_{2d}[J,l,h]={l \mathrm{dim} J \over l+h}+{\mathrm{dim} J \over 1+h}-{(l+1) \mathrm{dim} J \over l+1+h}.
\end{equation}
One can check the 2d central charge is related to 4d central charge as in formula \ref{eq:centralchargerelation}.
The Lie algebra data is shown in table \ref{table:liealgebra}.
\begin{table}[!htb]
\begin{center}
  \begin{tabular}{|c|c|c|c|}
    \hline
    J&$\mathrm{dim}(J)$&$\mathrm{rank}(J)$&h \\ \hline
    $A_{N-1}$&$N^2-1$&N-1&N\\ \hline
      $D_N$&$2N^2-N$& N& 2N-2 \\ \hline
       $E_6$&78&6&12 \\ \hline
              $E_7$&133&7&18 \\ \hline
       $E_8$&248&8&30 \\ \hline
       \end{tabular}
  \end{center}
  \caption{Lie algebra data.}
  \label{table:liealgebra}
\end{table}

We would like to make several comments,
\begin{itemize}
\item For our theory $J^b[k]$, the 2d chiral algebra is actually the minimal model of the W algebra and can be denoted as $W^J(p',p)=W^J(b+k,b)$  \cite{Bouwknegt:1992wg}. The 2d central charge 
takes the following form: 
\begin{equation}
c_{2d}= \text{rank}(J) \left[1-h(h+1){(p'-p)^2\over p p'}\right].
\end{equation}
\item To match the central charges of 4d and 2d, there are two choices of the levels $l,~l_1$ which satisfy the condition
\begin{equation} 
l+l_1=-(2h+1).
\end{equation}
These two levels have the property:  $l+h>0$ and $l_1+h<0$. The choice of $l$ makes the computation easier, and it would 
be interesting to consider other choice too.
\item For the $A_{N-1}^N[k]$ type theory, they can be either engineered using 6d $A_{N-1}$ theory
or 6d $A_{k-1}$ theory. This implies that the two cosets should be isomorphic by exchanging $k$ and $N$,
indeed such isomorphism has been confirmed in \cite{1991NuPhB.356..750K, Altschuler:1990th}.
\item The $D_4^6[3]$ theory is also called $(A_2, D_4)$ theory, and this theory is the same as the theory
considered in \cite{Buican:2016arp}. We have found a different realization of the chiral algebra from the one found in \cite{Buican:2016arp}. It would be interesting to
compare them.
\end{itemize}

Consider the chiral algebra ${\cal A}(J^h[k])=W^J[p',p]=W^J[h+k,h]$ with $(k,h)=1$.
the vacuum character takes the following simple form \cite{frenkel1992characters, Bouwknegt:1992wg}\footnote{For $J=A$, the character is already considered in \cite{Cordova:2015nma}.}:
\begin{equation}
\chi(q)={1\over \eta(\tau)^r}\sum_{\omega \in \hat{W}} \epsilon(w)q^{{1\over 2 p p'}|p'w(\Lambda^{+}+\rho)-p(\Lambda^{-}+\rho)|^2},
\end{equation}
here $r$ is the rank of the Lie algebra $J$, $\hat{W}$ is the affine Weyl group, and $\epsilon(w)$ is the signature of the affine Weyl group element.
$\Lambda^{+}\in P_{+}^{0}$ and $\Lambda^{-}\in P_{+}^{k}$ are principle admissible weights such that: 
\begin{equation}
{1\over 12}r h(h+1)(p'-p)^2=|p'(\Lambda^{+}+\rho)-p(\Lambda^{-}+\rho)|^2.
\end{equation}
We have the solution: $\Lambda^{+}=0$ and $\Lambda^{-}=(k,0,\ldots,0)$. Substitute into the character formula, we get the Schur index of our 4d theory $J^b[k]$. 
 Other cases with $b\neq h$ will be considered in a separate publication.

\section{Argyres-Douglas Matter}
Argyres-Douglas matters are defined as those \textbf{isolated} $\mathcal{N}=2$ SCFTs with following properties: a) The Coulomb branch operators
carry fractional scaling dimensions; b) they carry non-abelian flavor symmetries. They can be engineered by putting M5 branes on a sphere with
one irregular and one regular puncture. These theories can be labeled as $(J^{b}[k], Y)$ \cite{Wang:2015mra}.
The irregular singularity is labeled by $J^{b}[k]$ and the regular singularity is labeled by $Y$ \cite{Gaiotto:2009we, Chacaltana:2012zy}\footnote{When $b=h, J\neq A$, those 
theories are first studied in \cite{Cecotti:2012jx}.}. We focus on the theory whose irregular singularity does not have mass parameters.

Let us consider the simplest case where the regular puncture is full \cite{Chacaltana:2012zy}, then our theory can be labeled as
$(J^{b}[k], F)$. The 4d central charge  can be computed using the formula \ref{eq:acformula}, and we have the following observations:
\begin{equation}
u_{max}={1+(h-1)q(z)\over \sum q_i-1}=u_{0max}+{(h-1) q(z)\over \sum q_i-1};
\end{equation}
Here $q_i$ are the weights of the coordinates $x_i, z$ of the coordinates appearing in the 3-fold singularity of the irregular puncture. Other quantities appearing in the central charge formula are also easy to compute
\begin{equation}
r={\mu_0\over 2}+{\mathrm{dim}(J)-\mathrm{rank}(J)\over 2},~\mu=\mu_0+\mathrm{dim}(J).
\end{equation}

The flavor central charge is conjectured to be equal to the maximal
scaling dimension of the Coulomb branch operators:
\begin{equation}
k_F(J^b[k],F)=u_{max}=h-{b\over b+k},
\end{equation}
where $h$ is the dual Coxeter number of Lie algebra $J$.
The central charges of 4d theories are
listed in table. \ref{table:centralcharge}.  
\begin{table}[!htb]
\begin{center}
  \begin{tabular}{ |c|c|c|}
    \hline
       Theory &$c_{4d}$&$k_F$  \\ \hline
     $(A_{N-1}^N[k],F)$ &$\frac{1}{12}{(N+k-1)(N^2-1)}$& ${N(N+k-1)\over N+k}$\\ \hline
$(A_{N-1}^{N-1}[k],F)$&${(N+1)[N^2+N(k-2)+1]\over 12}$&${(N-1)^2+kN\over N+k-1}$\\ \hline

 $(D_N^{2N-2}[k],F)$   &${1\over 12}N(2N-1)(2N+k-3)$ & ${(2N-2)(2N+k-3)\over 2N-2+k}$ \\     \hline
  $(D_N^{N}[k],F)$  &${(2N-1)[2k(N-1)+N(2N-3)]\over 12}$ & ${2k(N-1)+N(2N-3)\over N+k} $ \\     \hline

 $(E_{6}^{12}[k],F)$ &${13(k+11)\over 2}$  &${12(k+11)\over k+12}$  \\     \hline
 $(E_{6}^{9}[k],F)$& ${13\over 6}(33+4k)$ &$12-{9\over k+9}$   \\     \hline
 $(E_{6}^{8}[k],F)$&${13\over 4}(22+3k)$ &$12-{8\over k+8}$   \\     \hline

$(E_{7}^{18}[k],F)$&${133\over 12}(17+k)$   &${18(k+17)\over k+18}$  \\     \hline
 $(E_7^{14}[k],F)$&${19\over 12}(119+9k)$  &$18-{14\over k+14}$     \\     \hline

 $(E_{8}^{30}[k],F)$ & ${62\over 3} (29+k)$  &${30(k+29)\over k+30}$  \\     \hline
      $(E_{8}^{24}[k],F)$ &${31\over 6} (116+5 k)$   & $30-{24\over k+24}$   \\     \hline
  $(E_{8}^{20}[k],F)$&${31\over3} (58+3 k)$  &$30-{20\over k+20}$ \\     \hline
  \end{tabular}
  \end{center}
  \caption{The central charge $c_{4d}$ and flavor central charge $k_F$ for Argyres-Douglas matter. }
  \label{table:centralcharge}
\end{table}

\subsection{2d chiral algebra}

We conjecture that the chiral algebra associated with the AD matter $(J^{b}[k], F)$ is the affine Kac-Moody algebra,
\begin{equation}
\CA=J_{-k_F}.
\end{equation}
One can easily check the 2d central charge and 4d central charge are related in the correct way as in equation \ref{eq:centralchargerelation}.
The 2d vacuum character, hence the 4d Schur index can be computed using the formula presented in \cite{Kac:1988qc} if $b=h,  (b,k)=1$, other cases 
would be considered elsewhere.

Now let us study some examples. Consider the simplest model $(A_{2N}^{2N+1}[-2N+1], F)=D_2[SU(2N+1)]$. The 2d chiral algebra is the Kac-Moody algebra $SU(2N+1)_{-{2N+1\over2}}$. For $D_2[SU(5)]$, the first few orders of vacuum character of the affine Kac-Moody algebra is 
\begin{equation}
\label{eq:SchurD2A4}
\CI_{D_2[SU(5)]}=\mathrm{PE}[(q+q^3+q^5+q^7+\cdots)\chi^{SU(5)}_{adj}],
\end{equation}
and for $D_2[SU(7)]$ we have
\begin{equation}
\CI_{D_2[SU(7)]}=\mathrm{PE}[(q+q^3+\cdots)\chi^{SU(7)}_{adj}],
\end{equation}
where $\chi^{\CG}_{adj}$ is the character of the adjoint representation of Lie algebra $\CG$ and $\mathrm{PE}$ means the plethystic exponential
\begin{equation}
\mathrm{PE}[x]=\exp\left[\sum_{n=1}^{\infty}\frac{1}{n}x^n\right]=\frac{1}{1-x}.
\end{equation}
We conjecture that the Schur index of $D_2[SU(2N+1)]$  is
\begin{equation}
\label{eq:schurindexD2A2N}
\CI_{D_2[SU(2N+1)]}=\mathrm{PE}\left[\frac{q}{1-q^2}\chi^{SU(2N+1)}_{adj}\right].
\end{equation}
Notice this is the same as the Schur index of half-hypers in adjoint representation of $SU(2N+1)$ with $q$ substituted by $q^2$ \cite{Gadde:2011ik,Gadde:2011uv}, or in 2d language the vacuum character of $SU(2N+1)_{-{2N+1\over 2}}$ is the partition function of free symplectic bosons in adjoint representation of $SU(2N+1)$ with $q$ replaced by $q^2$.

\begin{figure}
\centering
\begin{tikzpicture}[scale=0.5]
\node[W](1) at (0,0){1};
\node[W](2) at (1,-1){2};
\node[W](3) at (2.5,-2.5){N};
\node[W](4) at (4.5,-2.5){N};
\node[W](5) at (6,-1){2};
\node[W](6) at (7,0){1};
\node[W](7) at (3.5,-4.23){1};

\draw[-](1)--(2);
\draw[dotted](2)--(3);
\draw[-](3)--(4);
\draw[dotted](4)--(5);
\draw[-](6)--(5);
\draw[-](3)--(7);
\draw[-](4)--(7);
\end{tikzpicture}
\caption{\label{fig:quiver:D2A2N}The 3d mirror quiver of $D_2(SU(2N+1))$ theory.}
\end{figure}
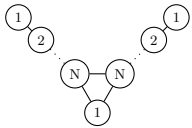

The Schur index of $D_2[SU(2N+1)]$ (equation \ref{eq:schurindexD2A2N}) also reproduces the correct chiral ring relations of the theory. One can check this by comparing with the Hall-Littlewood (HL) index of the same theory.  The HL index of the original 4d theory is the same as the Coulomb branch Hilbert series of
the mirror theory \cite{Gadde:2011uv,Cremonesi:2013lqa,Cremonesi:2014kwa}. If the mirror theory admits a quiver description, one can compute its Coulomb branch Hilbert series following the method in \cite{Cremonesi:2013lqa,Cremonesi:2014kwa}.

For example the quiver of the mirror theory of $D_2[SU(2N+1)]$ is shown in figure \ref{fig:quiver:D2A2N} \cite{Xie:2016uqq}. the HL index of $D_2[SU(5)]$ is
\begin{equation}
\mathrm{HL}_{D_2[SU(5)]}=\mathrm{PE}\left[\chi_{\mathbf{24}}t-(1+\chi_{\mathbf{24}})t^2+\chi_{\mathbf{24}}t^3+\dots\right],
\end{equation}
which means in the chiral ring of $D_2[SU(5)]$ there are dimension 2 and 6 generators in $\mathbf{24}$ (adjoint) representation of $SU(5)$ together with dimension 4 relations in $\mathbf{24}+\mathbf{1}$ representation and more. Exactly same generators and relations can also be extract from the Schur index \ref{eq:SchurD2A4} by noticing that the Schur indices of dimension $d$ chiral operator (relations) are $\frac{q^{d/2}}{1-q}$ ($-\frac{q^{d/2}}{1-q}$) instead of $t^{d/2}$ ($-t^{d/2}$) in HL index and stress tensor multiplet contributes $q^2\over 1-q$ to the Schur index but not present in HL index. One can also check that the Schur and HL indices of $D_2[SU(7)]$ produce the same  generators and relations of the chiral ring at lower dimension.

The Schur index can also be computed from the trace of the monodromy built from the BPS spectrum \cite{Cecotti:2010fi,Cordova:2015nma,Cecotti:2015lab}. For $D_2[SU(2N+1)]$ theory, the minimal chamber has $2N(2N+1)$ BPS states using the tools developed in \cite{Xie:2012dw,Xie:2012gd}. Unfortunately, 
for $D_2[SU(5)]$ theory, the minimal chamber has 20 states, we are not able to compute the trace of the monodromy due to limiting computing power. 

\section{Discussions}
The chiral algebra for theory considered in this paper is strikingly simple, and it is closed related
to the six dimensional construction, for example, the Lie algebra appearing in the coset model and
affine Kac-Moody algebra is precisely the type of 6d $(2,0)$ theory we start with. It would be interesting to interpret why this same
Lie algebra appears in 2d chiral algebra description. 

In this paper, we have considered only those irregular singularities without mass parameters.
For other type of irregular singularities,  one usually has a set of extra $U(1)^f$ flavor symmetries, and the 2d chiral algebra should  have extra generators corresponding 
to the moment map and other Higgs branch generators of $U(1)^f$ symmetries. The 4d central charge formula takes the following form 
\begin{equation}
12c_{4d}=-c_{2d}-f,
\end{equation}
here $c_{2d}$ is the central charge from the coset or the affine Kac-Moody part. The above observation motivates the following conjecture about the 
chiral algebra of the general case:   One has the same coset or affine Kac-Moody factor, and we add extra generator corresponding to Higgs branch operators of $U(1)^f$ factor, and 
the 2d central charge of this piece is $-f$. For example, $D_2[SU(2N)]$ is $SU(N)$ SQCD with $N_f=2N$, according to our proposal, the chiral algebra consists of a affine Kac-Moody  
factor $SU(2N)_{-N}$, and we should have extra generators corresponding to the extra $U(1)$ flavor symmetries. This structure agrees with the conjecture made in \cite{Beem:2013sza}.

We can gauge AD matter to form new $\mathcal{N}=2$ SCFT. Since we know the chiral algebra
of the AD matter and the gluing procedure of the chiral algebra, it would be interesting to find
the chiral algebra for theories built from gauging AD matter.
It would be also interesting to find out the chiral algebra of the general $\mathcal{N}=2$ SCFT engineered from
three-fold singularity \cite{Xie:2015rpa}.

The $J^b[k]$ theory can be derived by closing off the full puncture of $(J^b[k],F)$, and it is argued in \cite{Beem:2013sza} that the closing 
off puncture procedure corresponds to Drinfeld-Sokolov (DS) reduction for corresponding 2d chiral algebra. The chiral algebra of $J^b[k]$ theory is W algebra $W^J(b+k,k)$, and 
indeed it can be derived from DS reduction from the affine Kac-Moody algebra $J_{-h+b/(b+k)}$ \cite{frenkel1992characters} which is precisely the chiral algebra 
of the theory $(J^b[k],F)$ \footnote{We thank L.Rastelli for this very interesting observation.}.

%
\section*{Acknowledgments}
The work of S.T Yau is supported by  NSF grant  DMS-1159412, NSF grant PHY-
0937443, and NSF grant DMS-0804454.
The work of DX and WY is supported by Center for Mathematical Sciences and Applications at Harvard University. We would like to thank B. Lian, S. Cecotti, L. Rastelli, J. Song ,Y. Wang and especially C.Vafa for invaluable discussions.
%

\vspace{1cm}

\bibliographystyle{apsrev4-1}
\bibliography{refs1}


\end{document}